\documentclass{article}
\usepackage{amsmath}
\DeclareMathOperator{\rot}{rot}
\DeclareMathOperator{\diver}{div}
\numberwithin{equation}{section}
\newcommand{\bm}[1]{\mbox{\boldmath$#1$}}
\def\vo#1{{\bf #1}}
\def\no#1{, No.~#1}

\title{INVERSE PROBLEM FOR STATIC ELECTROMAGNETIC FIELD IN A DIPOLE
APPROXIMATION.}

\author{V.Ya. Epp$^*$, G.F. Kopytov$^{**}$ and T.G. Mitrofanova$^*$\\ 
$^*$Tomsk State Pedagogical University,\\
634041 Tomsk, Komsomolsky pr. 75, Russia.\\E-mail:
epp@tspu.edu.ru\\
$^{**}$Kuban State University, 350040 Krasnodar, Russia}

 \date{}
\begin{document}
\maketitle


\abstract{ The following inverse problem is discussed. A static 
electromagnetic field ge\-ne\-ra\-ted by a limited system of charges and 
currents is supposed to be known with its first derivatives at a point 
somewhere far from the system. This allows to reconstruct the position of the 
system, its net charge, and the electric and magnetic moments of the system.} 

\section{Statement of the problem.}

 By the inverse problem of
electrodynamics we mean the problem of reconstructing the
charge and current densities from the known elec\-tro\-mag\-ne\-tic
field they create. Obviously, the solution of the inverse problem of
electrodynamics depends substantially on the presence of materials, the
region in which the electromagnetic field is specified, ets. It
is well known that the value of an analytic function and all its
derivatives at some point allow to reconstruct the value of the
function in the domain of definition by use of the Taylor series.
Applied to the electrostatic inverse problem it means that if we
know the value of electromagnetic field in some point and all its
derivatives at this point, we can reconstruct the field in the whole
space.  Furthermore, the Maxwell equations allow to calculate the
charge and current density functions $\rho(\bm r)$ and $\bm j(\bm r)$
respectively:
$$\rho(\bm r)=\frac{1}{4\pi}\diver\bm E(\bm r)\;\;\;\;
\bm j(\bm r)=\frac{c}{4\pi}\rot\bm H(\bm r).$$
Here $\bm E(\bm r)$ is the electric field and $\bm H(\bm r)$ is the
magnetic field.

In actual practice we are able to measure the field and only few first
derivatives with an inevitable error. This means that we can
reconstruct the field only in a small vicinity around the point where
the field is me\-a\-su\-red. But the real problem encountered in
practice is to calculate the field and charge distribution far from the
point where the field is measured. In order to do it we have to know
the value of field and its derivatives up to a very high order and with
high accuracy.  We present here another approach based on the multipole
expansion of the field far from the static collection of charges and
currents.

Suppose we know the value of electric $\bm E$ and magnetic $\bm H$
fields at some point far from limited system of charges and
currents. We know also the first derivatives of fields $E_{ij}=\partial
E_i/ \partial x_j$ and $H_{ij}=\partial H_i/ \partial x_j$. There is to
define the charge $q$, the electric $\bm d$ and magnetic $\bm m$ dipole
moments of the distant electromagnetic system, and the location $\bm r$
of the system. Let the coordinate origin be at the point where the
fields are defined. We start with the well known formula for the fields
of electric and magnetic dipoles [\ref{Land}, \ref{Grif}]
\begin{eqnarray}\label{in1}
\bm E&=&-\frac{q\bm r}{r^3}+\frac{3(\bm r\bm d)\bm r-r^2\bm d}{r^5}\\
\label{in1a}
\bm H&=&\frac{3(\bm r\bm m)\bm r-r^2\bm m}{r^5}
\end{eqnarray}
Taking the derivatives of these expressions with respect to coordinates
we find
\begin{eqnarray}\label{in2}
E_{ij}&=&\frac{\partial E_i}{\partial
x_j}=\delta_{ij}\left[-\frac{q}{r^3}+ 3\frac{\bm r\bm
d}{r^5}\right]+3\frac{d_jr_i+d_ir_j}{r^5}\nonumber\\&+&3r_ir_j
\left[\frac{q}{r^5}-5\frac{\bm r\bm d}{r^7}\right]\\
\label{in2a}
H_{ij}&=&\frac{\partial H_i}{\partial x_j}=\delta_{ij}3\frac{\bm r\bm
m}{r^5}+3\frac{m_jr_i+m_ir_j}{r^5}-15r_ir_j\frac{\bm r\bm m}{r^7}
\end{eqnarray}
where $\delta_{ij}$ is the Kronecker symbol. One can see from Eqs.
(\ref{in2}) and (\ref{in2a}) that the tensors $E_{ij}$ and $H_{ij}$ are
symmetrical. This follows also from the Maxwell equations, namely from
$\rot\bm E=0$ and $\rot\bm H=0$. Besides,
the Maxwell equations $\diver\bm E=0$ and $\diver\bm H=0$
give
\begin{eqnarray}\label{2}
E_{11}+E_{22}+E_{33}=0
\quad
H_{11}+H_{22}+H_{33}=0
\end{eqnarray}
We consider expressions (\ref{in1}) -- (\ref{in2a}) as a set of
equations in 10 scalar unknowns $q$, $\bm r$, $\bm d$ and $\bm m$.
But this set consist of 16 equations with regard to the conditions
(\ref{2}), and symmetry of the tensors $E_{ij}$ and
$H_{ij}$. This allows us to formulate the inverse problem in more
general sence. Namely, we can consider fields $\bm E$ and $\bm H$ as
being originated by different sources. Let us denote by $\bm r_q$ the
radius-vector of a charge collection with a total charge $q$ and an
electric dipole moment $\bm d$, and by $\bm r_m$ the radius-vector of a
system of currents with a magnetic dipole moment $\bm m$. Then the set
of equations (\ref{in1}) -- (\ref{in2a}) splits into two independent
sets for $\bm r_q$ and $\bm r_m$. In solving these equations we can
obtain in particular $\bm r_q=\bm r_m$, which means that the electric
field $\bm E$ and magnetic field $\bm H$ are generated by the same
electromagnetic system.

It should be pointed out, that in calculation of fields far from an
electromagnetic system we can neglect the higher terms of multipole
expansion. The accuracy of such representation depends on the ratio
between the sizes of the system and the distance between the system and
observer. In solving the inverse problem we find only the distance
$|\bm r|$, but not the sizes of system. In order to estimate the
accuracy of the received solution one have to calculate the sizes of
the system by some independent method. For example one can solve the
inverse problem for a few different positions of the observer and then
one can estimate the sizes of the system. One can also investigate the
field in the vicinity of the observer in order to find out whether this
field is of dipole nature. These problems are not considered in this
paper. In order to exclude such questions we consider further a
point-like charge with an electric dipole moment and a point-like magnetic
moment. In the next two sections we solve the inverse problem first for
the equations (\ref{in1}) and (\ref{in2}) and then for equations
(\ref{in1a}) and (\ref{in2a}).

\section{Inverse problem for a charge and electric dipole.}
Let us solve the equations (\ref{in1}) and (\ref{in2}) with respect to
the unknown $q$, $\bm d$ and $\bm r$\footnote{We omit the subscripts
$q$ and $m$ having in mind that in the section 2 we are dealing only
with $\bm r_q$ and in section 3 only with $\bm r_m$}.  We choose the
coordinate system as follows:  the coordinate origin is placed at the
point where the field is specified, the $x$ axis is directed along the
vector $\bm E$ and the $y$ axis is aligned with the principal normal to
the electric field line.  The vector of the principal normal $\bm n$ is
defined by the equality [\ref{Korn}]
\begin{eqnarray} \bm
 n=\frac{1}{kE}\frac{\partial\bm E}{\partial s}\nonumber
\end{eqnarray}
where $k$ is the curvature of the field line and $\partial s$ is the
displacement along the field line. Taking derivative from $\bm E$ we
find
$$\bm n=\frac{1}{kE^4}[\bm E[\bm D\bm E]]$$
where $\bm D=(\bm E\bm\nabla)\bm E$ with the projections
$D_i=E_kE_{ik}$ (a summation over repeated indices is implied). The
projections of the vector $\bm n$ on the axes of an arbitrary
orthogonal coordinate system are expressed in terms of components of
the tensor $E_{ij}$ as follows:
\begin{eqnarray}
n_i=\frac{1}{kE^4}E_kE_j(E_jE_{ik}-E_iE_{kj}).\nonumber
\end{eqnarray}
If the $x$ axis is directed in the sense of the vector $\bm E$, then
\begin{eqnarray}
\bm n=\frac{1}{kE}\{0,\;E_{12},\;E_{13}\}.
\end{eqnarray}
With the $y$ axis directed in the sense of the vector $\bm n$ we have
$E_{13}=0$ and $E_{12}>0$. It can be seen from Eq. (\ref{in1}) that the
vectors $\bm E$, $\bm d$ and $\bm r$ lie in the same plane; therefore,
the problem is reduced to a two-dimensional one. In the $x$, $y$, $z$
coordinate system we have $\bm E=(E,0,0)$, $\bm d=(d_x,d_y,0)$, and
$E_{23}=0$ as a consequence of the axial symmetry of the field.
Thus, five functions of the coordinates $E_{x}$, $E_{y}$, $E_{11}$,
$E_{12}$ and $E_{33}$ are known. They can be used to find five unknowns
$q$, $d_x$, $d_y$, $r_1$ and $r_2$. Eqs.(\ref{in1}) and (\ref{in2}) in
 the $(x,y)$ plane take the form
\begin{eqnarray}\label{in3a}
E&=&-\frac{qx}{r^3}+\frac{3(\bm r\bm d)x-r^2d_x}{r^5}\\
\label{in3b}
0&=&-\frac{qy}{r^3}+\frac{3(\bm r\bm d)y-r^2d_y}{r^5}\\
\label{in3c}
E_{11}&=&\frac{q}{r^3}-3\frac{(\bm r\bm d)}{r^5}-6\frac{d_xx}{r^5}
-3x^2\left[\frac{q}{r^5}-5\frac{(\bm r\bm d)}{r^7}\right]\\
\label{in3d}
E_{12}&=&-3\frac{d_xy+d_yx}{r^5}-3xy\left[\frac{q}{r^5}-5\frac{(\bm r
\bm d)}{r^7}\right]\\
\label{in3e}
E_{33}&=&\frac{q}{r^3}-3\frac{(\bm r\bm d)}{r^5}.
\end{eqnarray}
The charge $q$ and the components of the dipole moment can be
easily expressed from Eqs. (\ref{in3a}), (\ref{in3b}) and (\ref{in3e}):
\begin{eqnarray}\label{in5}
d_x&=&-r^3(E+E_{33}x)\\
d_y&=&-r^3yE_{33}\\
\label{in14}
q&=&-2r^3E_{33}-3xrE
\end{eqnarray}
Substituting these formulas into Eqs. (\ref{in3c}) and (\ref{in3d}), we
find
\begin{eqnarray}\label{in6}
E_{11}&=&E_{33}+6\frac{xy^2E}{r^4}-3\frac{x^2E_{33}}{r^2}\nonumber\\
E_{12}&=&-3\frac{xyE_{33}}{r^2}+3\frac{yE}{r^2}-6\frac{yx^2E}{r^4}.
\end{eqnarray}
Taking into account that $r=\sqrt{x^2+y^2}$, we get two equations
depending only on $x$ and $y$. Thus, the problem is reduced to  the
solution of the last system of equations (\ref{in6}). By simple
algebraic manipulations this system can be reduced to the following
form $(r\neq0)$:
\begin{eqnarray}\label{in7}
x^2(2E_{33}+E_{11})+y^2(E_{33}-E_{11})+2yxE_{12}&=&0\nonumber\\
-xy(E_{33}+2E_{11})-y^2E_{12}+x^2E_{12}+3yE&=&0.
\end{eqnarray}
Let us introduce the designations
\begin{eqnarray}
F_1&=&E_{22}-E_{33},\;\;\;F_2=E_{33}-E_{11},\;\;\;F_3=E_{11}-E_{22},
\nonumber\\ S&=&\sqrt{E_{12}^2+F_1F_2}.\nonumber
\end{eqnarray}
Then the solution of Eq. (\ref{in7}) can be written in the form
\begin{eqnarray}\label{in8}
x=\frac{3E}{4E_{12}^2+F_3^2}\left[F_3\pm\frac{E_{12}(F_1-F_2)}{S}\right]
 \end{eqnarray}
\begin{eqnarray}\label{in9}
y=\frac{3E}{4E_{12}^2+F_3^2}\left[2E_{12}\pm\frac{2E_{12}^2-F_1F_3}{S}
\right].
\end{eqnarray}
In order to calculate the dipole moment, we find the squared
radius-vector
$r$.
\begin{eqnarray}\label{in10}
r^2=\frac{9E^2}{S^2(4E_{12}^2+F_3^2)}\left[2E_{12}^2-F_1F_3
\pm2SE_{12}\right]
\end{eqnarray}
Substituting Eqs. (\ref{in8}), (\ref{in9}) and (\ref{in10}) into Eq.
(\ref{in5}), we find the dipole moment
\begin{eqnarray}\label{in12}
d_x&=&\mp E^{10}(E_{12}(F_1^2+(-E_{12}\pm S)^2)\mp
2F_1^2S)\nonumber\\
&\times&\frac{27F_1^3}{(S)^4(F_1^2+(-E_{12}\pm S)^2)^{5/2}}\\
\label{in11}
d_y&=&\pm\frac{81E^4F_1^5E_{33}}{(S)^4
(F_1^2+(-E_{12}\pm S)^2)^{5/2}}
\end{eqnarray}
The sign $\pm$ in the formulas indicates the existence of two solutions
of the initial system of equations (\ref{in3a}-\ref{in3e}). Physically
this means that the same field with its derivatives may be created at
the given point by two different sources located at different places.
One can find some specific cases of this solution in Ref. [\ref{Epp}].

\section{Inverse problem for a magnetic dipole moment.}

In order to find the position vector $\bm r$ and the magnetic moment
 $\bm m$ of a particle generating the field $\bm H$ we solve equations
(\ref{in1a}) and (\ref{in2a}). Now we align the axis $x$ along the
vector $\bm H$ and the axis $y$ along the principal normal to the
magnetic field line. Repeating the reasoning of the previous section we
get
\begin{eqnarray}\label{in14a}
H&=&\frac{3(\bm r\bm m)x-r^2m_x}{r^5}\\
\label{in14b}
0&=&\frac{3(\bm r\bm m)y-r^2m_y}{r^5}\\
\label{in14c}
H_{11}&=&-3\frac{(\bm r\bm
m)}{r^5}-6\frac{m_xx}{r^5} +15x^2\frac{(\bm r\bm m)}{r^7}\\
\label{in14d}
H_{12}&=&-3\frac{m_xy+m_yx}{r^5}+15xy\frac{(\bm r\bm m)}{r^7}\\
\label{in14e}
H_{33}&=&-3\frac{(\bm r\bm m)}{r^5}
\end{eqnarray}

Thus, we have 5 equations for 4 unknowns. Hence, the system
of equations (\ref{in14a}) -- (\ref{in14e}) is a overdetermined one. It
means that the components of $\bm H$ and $H_{ij}$ are not independent.
The fact that the field $\bm H$ is produced by a magnetic moment places
a constraint on $\bm H$ and $H_{ij}$. The corresponding equation will
be found latter (see Eq. (\ref{in19})). Substituting $(\bm r\bm m)$
from Eq. (\ref{in14e}) into Eqs. (\ref{in14a}) -- (\ref{in14d}), we
find
\begin{eqnarray}\label{in15a}
H&=&-H_{33}x-\frac{m_x}{r^3}\\
\label{in15b}
 0&=&-H_{33}y-\frac{m_y}{r^3}\\
\label{in15c}
H_{11}&=&H_{33}-6\frac{m_xx}{r^5}-5x^2\frac{H_{33}}{r^2}\\
\label{in15d}
H_{12}&=&-3\frac{m_xy+m_yx}{r^5}-5xy\frac{H_{33}}{r^2}
\end{eqnarray}
Eliminating $m_x$ and $m_y$ we get
\begin{eqnarray}\label{in16a}
H_{11}&=&H_{33}+x^2\frac{H_{33}}{r^2}-6x\frac{H}{r^2}\\
\label{in16b}
H_{12}&=&3\frac{H}{r^2}y+xy\frac{H_{33}}{r^2}
\end{eqnarray}
and for the square of the distance
\begin{eqnarray}\label{in17}
r^2=-\frac{3Hx}{2H_{33}}.
\end{eqnarray}
We suppose here that $H_{33}\neq 0$. The case $H_{33}=0$ will be
considered latter. Using Eqs. (\ref{in16a}) -- (\ref{in16b}) one can
find
\begin{eqnarray}\label{in18a}
x=-\frac{3H(3H_{33}+H_{11})}{2H_{33}^2}
\end{eqnarray}
\begin{eqnarray}\label{in18b}
y=-\frac{3HH_{12}(3H_{33}+H_{11})}{2H_{33}^2(H_{11}+5H_{33})}
\end{eqnarray}
Now we express $r$ from Eq. (\ref{in17})
$$|r|=\frac{3|H|}{2|H_{33}|}\sqrt{\frac{3H_{33}+H_{11}}{H_{33}}}$$
And hence
\begin{eqnarray}\label{in18c}
m_x&=&\frac{1}{3}\left(\frac{3H}{2H_{33}}\right)^4(7H_{33}+3H_{11})
\left(\frac{3H_{33}+H_{11}}{H_{33}}\right)^{3/2}\\ \label{18c}
m_y&=&H_{12}\left(\frac{3H}{2H_{33}}\right)^4
\frac{3H_{33}+H_{11}}{H_{11}+5H_{33}}
\left(\frac{3H_{33}+H_{11}}{H_{33}}\right)^{3/2}
\end{eqnarray}

Let us consider the specific case $H_{33}=0$. It follows
 from Eq. (\ref{in14e}) that $(\bm r\bm m)=0$, which gives immediately
$x=0$. Eqs. (\ref{in14a}) -- (\ref{in14d}) give the whole solution in
this case
\begin{eqnarray}\label{*}
x=0\;\;\;y=3\frac{H}{H_{12}}\;\;\;m_x=-27\frac{H^4}{H_{12}^3}\;\;\;
m_y=0\end{eqnarray}
Thus, Eqs. (\ref{in18a}) -- (\ref{18c}) and in particular case
Eq. (\ref{*}) give the solution of the problem.

It was mentioned above that the system of Eqs. (\ref{in14a})
-- (\ref{in14e}) is a overdetermined one. Let us find relation between
$\bm H$ and $H_{ij}$. On the one hand it follows from Eq. (\ref{in17})
that
$$r^2=\frac{9H^2(3H_{33}+H_{11})}{4H_{33}^3},\;\;\;\;(H_{33}\neq 0).$$
On the other hand relations (\ref{in18a}) and (\ref{in18b}) give
$$r^2=\frac{9H^2(3H_{33}+H_{11})^2}{4H_{33}^4}\left(1+
\frac{H_{12}^2}{H_{33}^4}\frac{1}{(H_{11}+5H_{33})^2}\right).$$
This gives the desired equation
\begin{eqnarray}\label{in19}
\frac{H_{33}}{3H_{33}+H_{11}}=1+\frac{H_{12}^2}{(H_{11}+5H_{33})^2}.
\end{eqnarray}
In case $H_{33}=0$ we have from Eqs. (\ref{in14a}) -- (\ref{in14e})
$H_{11}=0$.

If the condition (\ref{in19}) is not fulfilled, the field $\bm H$ is
not produced by a magnetic dipole.

\vspace{1mm}
{\bf References}

\begin{enumerate}
\item\label{Land}
 L.D. Landau and E.M. Lifshitz, {\sl The Classical Theory of Fields.}
Pergamon, New York, 1975.
\item\label{Grif}
D.J. Griffiths, {\sl Introduction to Electrodynamic.}
Prentice Hall, N.Jer\-sy, 1999.
\item\label{Korn}
G.A. Korn and T.M. Korn, {\sl Mathematical Handbook.} McGraw-Hill, New
York, 1968.
\item\label{Epp}
V.Ya. Epp and T.G. Mitrofanova, Inverse problem for static
elec\-tro\-mag\-ne\-tic field. {\sl Russian Phys. J.} (1999)
\vo{42}\no{7}, 587--591.  \end{enumerate}

 \end{document}